\documentclass[12pt]{article}

\usepackage{amsmath,amssymb,amsfonts,amsbsy}
\usepackage{cite}
\usepackage{graphicx}
\usepackage{wrapfig}
\usepackage{epsfig}

\usepackage{bbm} 
\usepackage{bm} 
\usepackage{color}                                                       %
\usepackage{dsfont} 
\usepackage{latexsym} 
\usepackage{lscape} 
\usepackage{mathrsfs} 
\usepackage{morefloats} 
\usepackage{floatflt} 
\usepackage{slashed} 
\usepackage{psfrag}

\DeclareFontFamily{OT1}{mygreek}{}%
\DeclareFontShape{OT1}{mygreek}{m}{n}{<->omsegr}{}%
\DeclareFontShape{OT1}{mygreek}{b}{n}{<->omsegrb}{}%
\DeclareFontShape{OT1}{mygreek}{m}{it}{<->omsegri}{}%
\DeclareFontShape{OT1}{mygreek}{bx}{n}{<->sub * mygreek/b/n}{}%
\DeclareFontShape{OT1}{mygreek}{m}{sl}{<->sub * mygreek/m/it}{}%
\DeclareSymbolFont{Greekrm}{OT1}{mygreek}{m}{n} 
\DeclareSymbolFont{Greekbf}{OT1}{mygreek}{b}{n} 
\DeclareSymbolFont{Greekit}{OT1}{mygreek}{m}{it} 
\DeclareMathSymbol{\omegab}{\mathalpha}{Greekbf}{119}

\begin{document}
\addcontentsline{toc}{subsection}{{Longitudinal double spin asymmetry $A_1^p$ and spin-dependent structure function $g_1^p$ of the proton at low $x$ and low $Q^2$ from COMPASS}\\
{\it A.S. Nunes}}

\setcounter{section}{0}
\setcounter{subsection}{0}
\setcounter{equation}{0}
\setcounter{figure}{0}
\setcounter{footnote}{0}
\setcounter{table}{0}

\begin{center}
\textbf{LONGITUDINAL DOUBLE SPIN ASYMMETRY $A_1^p$\\ AND SPIN-DEPENDENT STRUCTURE FUNCTION $g_1^p$\\ OF THE PROTON AT LOW $x$ AND LOW $Q^2$ FROM COMPASS}

\vspace{5mm}

\underline{A.S.~Nunes}$^{\,1\dag}$, on behalf of the COMPASS Collaboration

\vspace{5mm}

\begin{small}
  (1) \emph{LIP, 1000-149 Lisbon, Portugal} \\
  $\dag$ \emph{E-mail: Ana.Sofia.Nunes@cern.ch}
\end{small}
\end{center}

\vspace{0.0mm} 

\begin{abstract}
The COMPASS experiment at CERN has collected a large sample of events of inelastic scattering of longitudinally polarised muons
off longitudinally polarised protons in the non-perturbative region 
(four-momentum transfer squared $Q^2<1$~(GeV$^2$/$c^2$), 
with 
a Bjorken scaling variable in the range 
$4\times 10^{-5}<x<4\times 10^{-2}$. The data set
is two orders of magnitude larger than the similar sample collected by the SMC experiment. These data
complement our data for polarised deuterons. They allow the accurate determination of the longitudinal double spin asymmetry
$A_1^p$ and of the spin-dependent structure function $g_1^p$
of the proton 
in the region of low $x$ and low $Q^2$. The preliminary results of the analysis of these data yield non zero and positive asymmetries and of the structure function $g_1^p$. This is the first time that spin effects are observed at such low $x$.
\end{abstract}

\vspace{7.2mm} 

In processes of inelastic scattering of leptons off nucleons, the region of low values of $x$ corresponds to high parton densities. Among experiments with polarisation, only fixed target experiments have been able, up to now, to probe that poorly known region. In this kind of experiments there is a strong correlation between $x$ and $Q^2$, which makes low $x$ measurements also low $Q^2$ ones, for which perturbative QCD is not valid. However, there are models that allow a smooth extrapolation to the low $Q^2$ region, while matching the perturbative QCD behaviour at high $Q^2$, including resummation or vector meson dominance \cite{BK,EGT}. The SMC experiment at CERN has done pioneer measurements of longitudinal double spin asymmetries $A_1^p$ and of the spin-dependent structure function $g_1^p$ of the proton down to $x=6\times 10^{-5}$ \cite{SMClowx}, but the measurements had a limited precision, that COMPASS can now improve. Furthermore, the non-singlet structure function $g_1^\text{NS}=g_1^p-g_1^n$, which decouples from gluons, can be extracted with increased precision when combining our published  and structure function $g_1^d$ of the deuteron at low $x$~\cite{COMPASSdeuteronQ2lt1} with these new preliminary results of $g_1^p$.

The COMPASS experiment is described in detain in \cite{COMPASSnima}. It is a fixed target experiment at the SPS using a tertiary, naturally polarised, muon beam. It consists of a large acceptance two-staged spectrometer with trackers and calorimetry in its two stages, and a RICH detector. A beam momentum of 160 GeV/$c$ was used in 2007, changed to 200 GeV/$c$ in 2011 to allow reaching lower values of $x$ for a given $Q^2$. The 1.2 meter-long polarised target was divided in three independent cells, to allow simultaneous data taking in two opposite spin configurations of the target material. In 2007 and 2011, the target consisted on polarisable protons from ammonia (NH$_3$). The polarisation of the target is built by the process of dynamic nuclear polarisation, using a superconducting solenoid with 2.5 T; in the frozen spin mode, the target material is kept at temperatures down to 50 mK using a $^3$He/$^4$He dilution refrigerator. Typical values of polarisation obtained were of the order of 85\%; the dilution factor, {\it i.e.} the percentage of polarisable material, was around  16\%.

\begin{figure}[h]
\centering
\begin{tabular}{cc}
\includegraphics[width=0.5\textwidth]{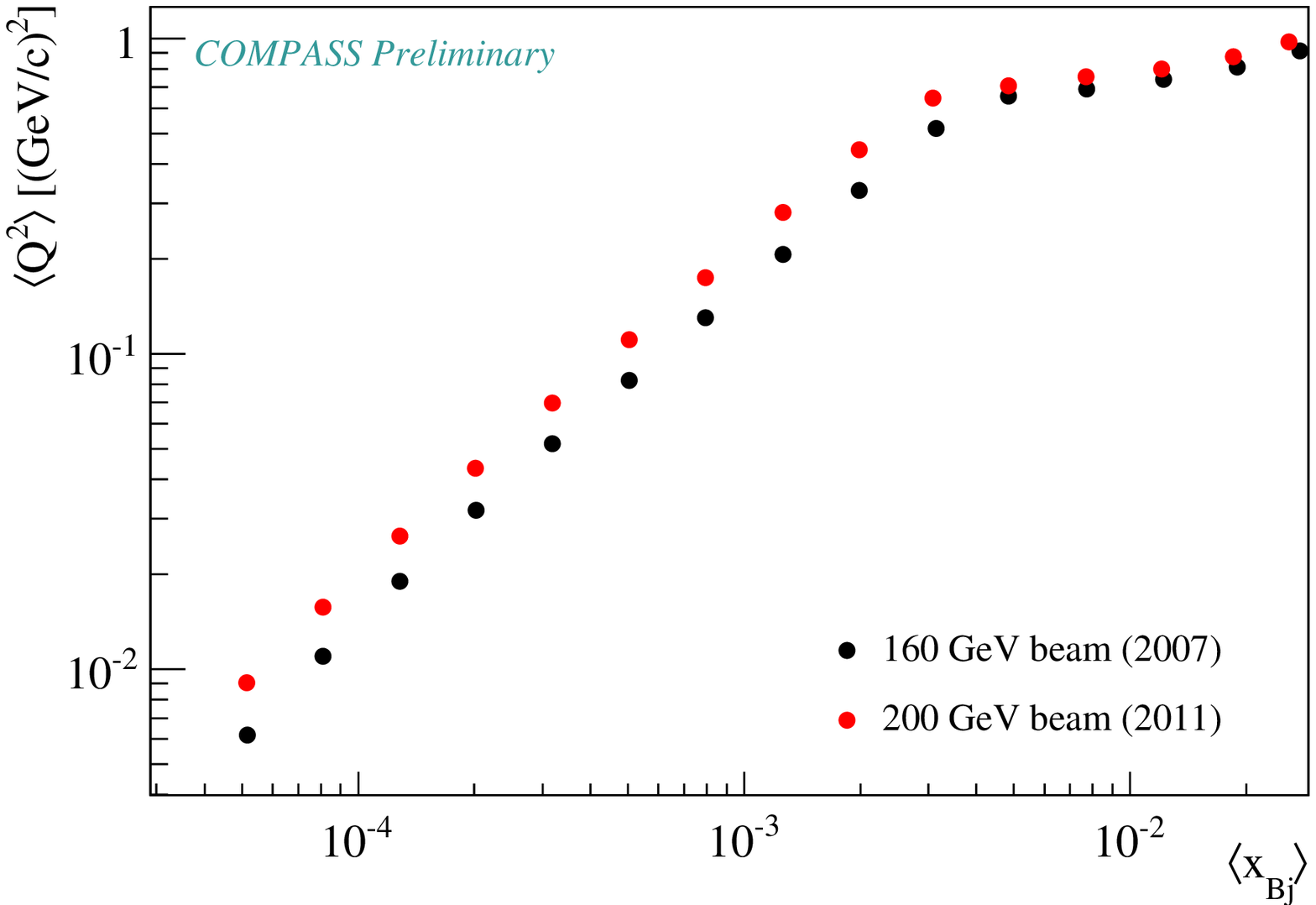}&
\includegraphics[width=0.5\textwidth]{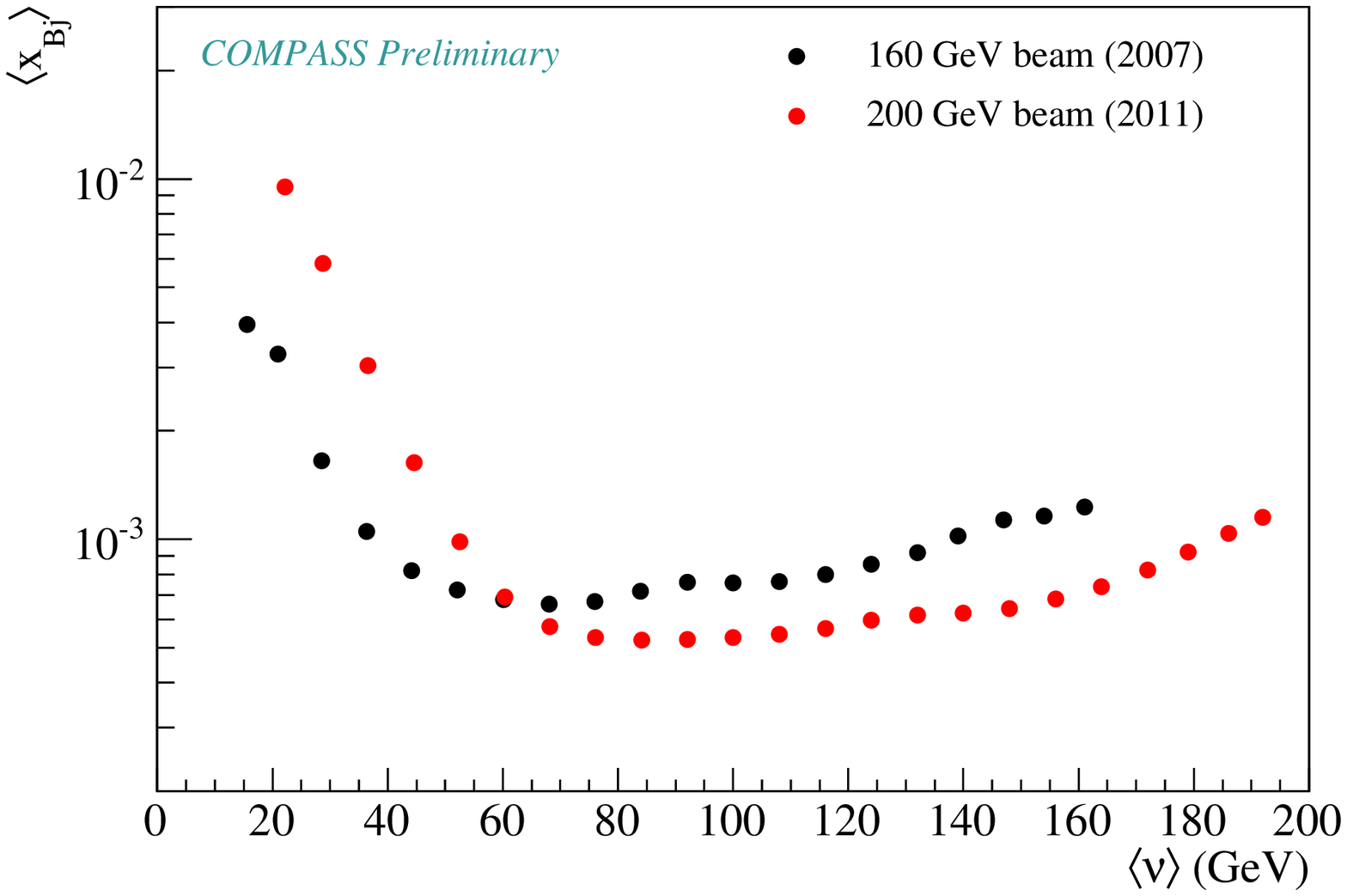}\\
    \textbf{(a)} & \textbf{(b)}\\
    \end{tabular}
      \caption{\footnotesize
    \textbf{(a)} Average values of $Q^2$ versus average values of $x$ of the bins used in the $A_1^p(x)$ and $g_1^p(x)$ extractions, for the 2007 and 2011 data samples. 
    \textbf{(b)} Similar plot, but for average values of $x$ versus average values of $\nu$ of the bins used in the $A_1^p(\nu)$ and $g_1^p(\nu)$ extractions. 
  }

  \label{fig:kinevars3}
\end{figure}

For the present analysis, data taken in 2007 and 2011 with a longitudinally polarised target of protons (from NH$_3$ divided in three cells, with neighbouring cells with opposite polarisation) and a longitudinally polarised muon beam with 160 or 200 GeV, respectively, were used. The main selection criteria of events were: 
(a) $Q^2<1$ (GeV/$c$)$^{2}$; 
(b) $x\geq 4\times 10^{-5}$; 
(c) 
the fraction of the muon energy transferred to the proton, in the laboratory, satisfied 
$0.1<y<0.9$;
(d) there was at least one additional track besides the scattered muon in the interaction vertex (to better discriminate the target cell in which the interaction occurred);
(e) the event was not due to elastic scattering of a muon off a target electron (which is a important contamination of the sample, peaked at $x=m_e/m_p\sim 5.5\times 10^{-4}$), as in~\cite{COMPASSdeuteronQ2lt1}.

The number of events in the two final samples with different beam energies are, respectively, $447\times 10^6$ and $229\times 10^6$, {\it i.e.} the COMPASS sample is about 150 times larger than the SMC one.
The average values, for the bins used in the analysis, of selected kinematic variables in the final samples are presented in Fig.~\ref{fig:kinevars3}.

The number of events in the antiparallel (parallel) spin configurations are given by:
$$N^{\stackrel{\leftarrow}{\Rightarrow},\stackrel{\leftarrow}{\Leftarrow}} \simeq a\phi n\bar{\sigma}(1\pm P_\text{beam}P_\text{target}fA_1^p )$$

The longitudinal double spin asymmetries of the proton, $A_1^p$, were extracted from data using a method~\cite{IvanovDspin13,COMPASSprotonQ2gt1} that weights each event by a factor $\omega=fDP_b$, {\it i.e.} the product of the dilution factor, the depolarisation factor and the polarisation of the beam, in order to reduce the statistical errors. 
Great care was taken to minimize possible sources of false asymmetries. This was done, on one hand, by only selecting events for which 
the beam track extrapolation crosses all the target cells, in order to have the same beam flux in all cells; and, on the other hand by using three target cells, by reversing the spin configuration of the target cells about every 24 hours, by measuring asymmetries independently in periods of data taken in about 48 hours, and by changing the combination of the target field and the spin configuration of the target cells at least once per year of data taking, to minimize the respective correlation.  

\begin{figure}[h]
\centering
\begin{tabular}{cc}
\includegraphics[width=0.5\textwidth]{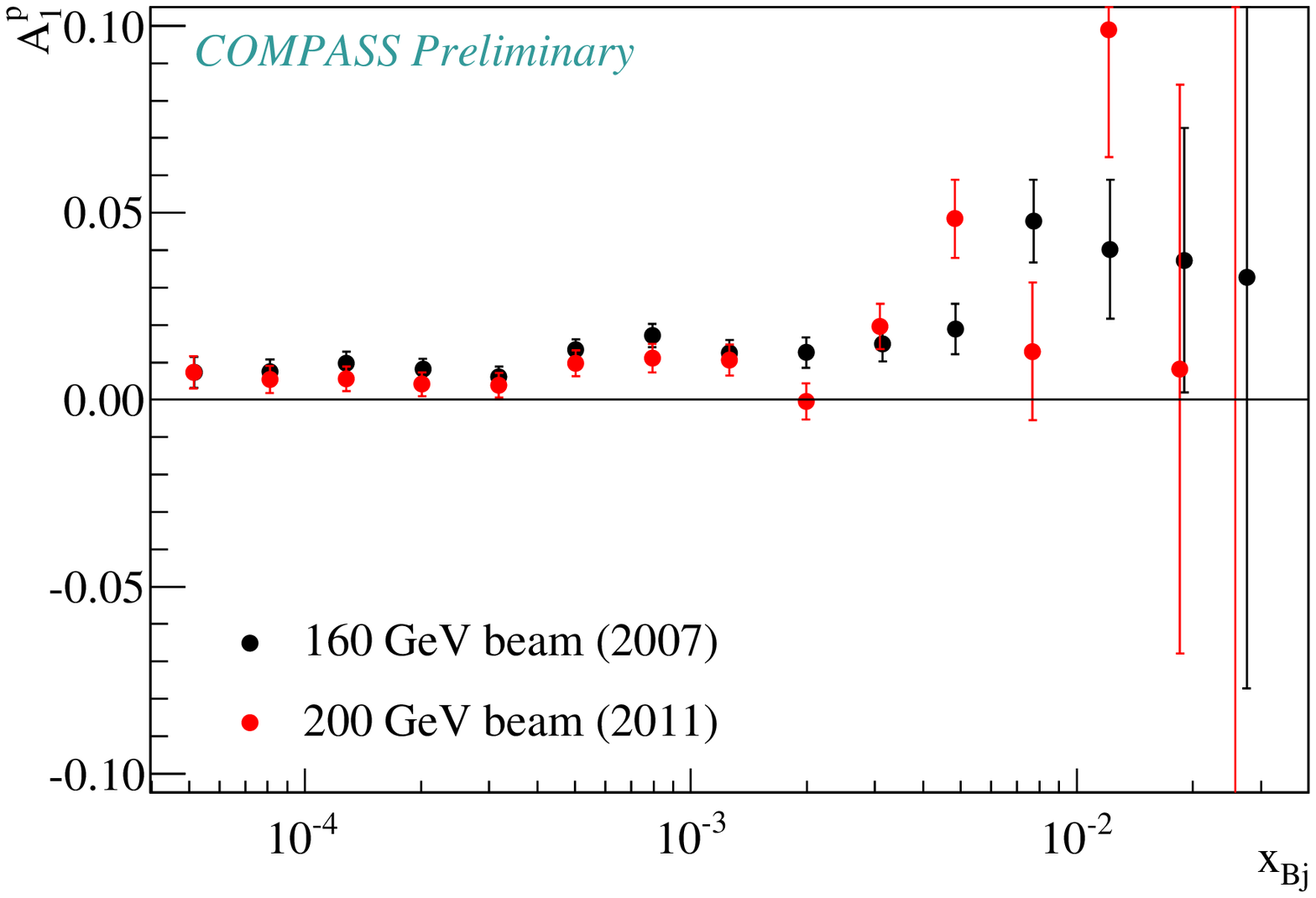}&
\includegraphics[width=0.5\textwidth]{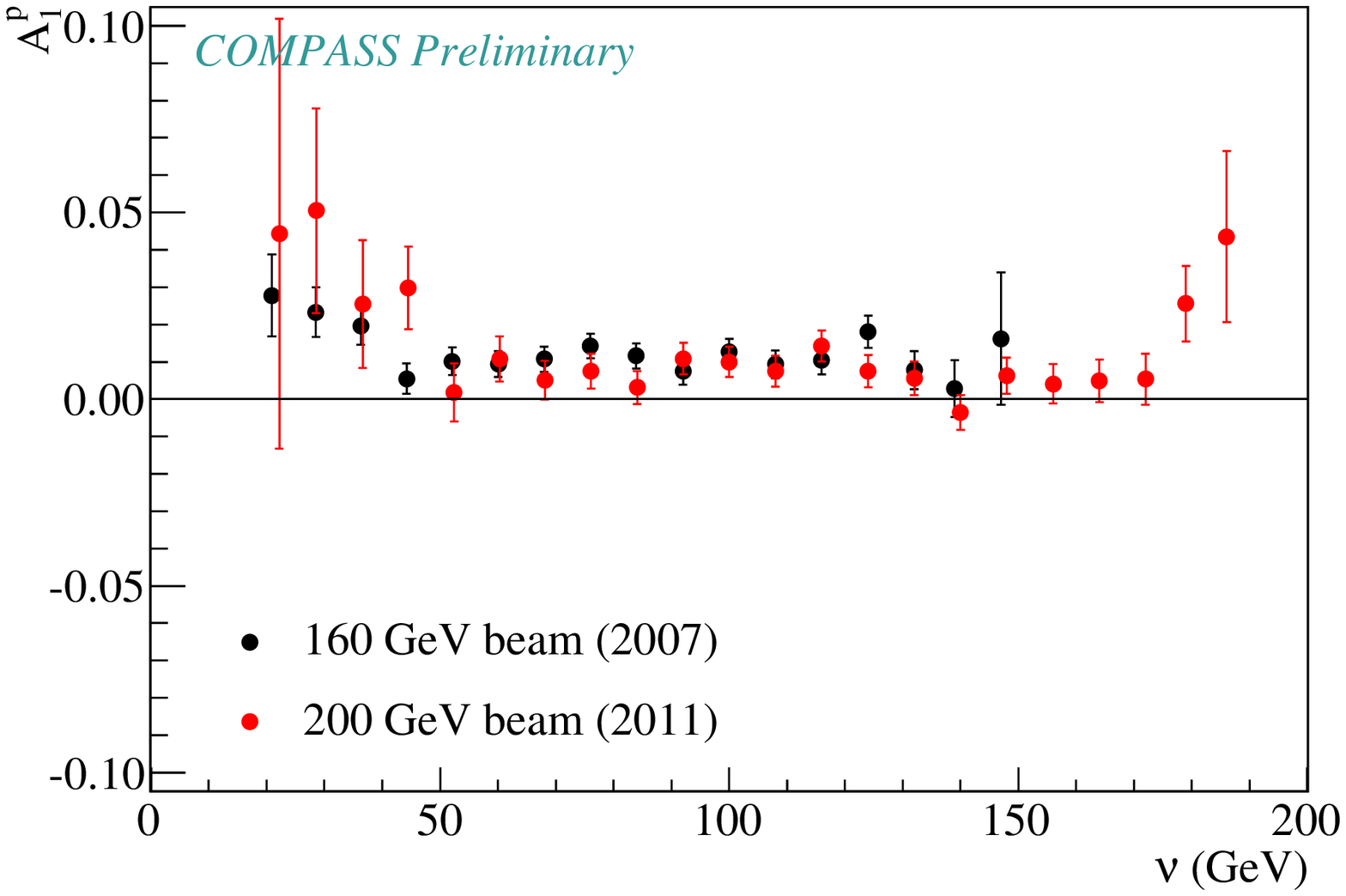}\\
    \textbf{(a)} & \textbf{(b)} \\
    \end{tabular}
      \caption{\footnotesize
    \textbf{(a)} Longitudinal double spin asymmetries $A_1^p$ as a function of $x$, obtained from 2007 and 2011 data, after corrections due to the polarised radiative asymmetry and the presence of $^{14}$N in the ammonia target. The results for the two beam energies are compatible within errors. The systematic errors are expected to be smaller than the statistical errors. The asymmetries are mostly incompatible with zero and positive. 
    \textbf{(b)} The same, but for $A_1^p$ as a function of $\nu$. 
  }
  \label{fig:A1}
\end{figure}

The asymmetries were obtained independently in bins of $x$ and in bins of the virtual photon energy $\nu$. Unpolarised radiative corrections taken from the program TERAD~\cite{TERAD} were included in an effective dilution factor, whereas polarised radiative corrections were taken from the program POLRAD~\cite{POLRAD}, and are less or equal than $25\%$ of the statistical errors. The asymmetries were further corrected for the presence of polarisable $^{14}$N in the target material, the corrections being less or equal than $1\%$ of the statistical errors. Thorough checks on possible sources of false asymmetries (which dominate the systematic errors) were done, and the systematic errors are expected to be smaller 
than the statistical errors. The preliminary results obtained for $A_1^p$ are shown in Fig.~\ref{fig:A1}. 

\begin{figure}[h]
\centering
\begin{tabular}{cc}
\includegraphics[width=0.5\textwidth]{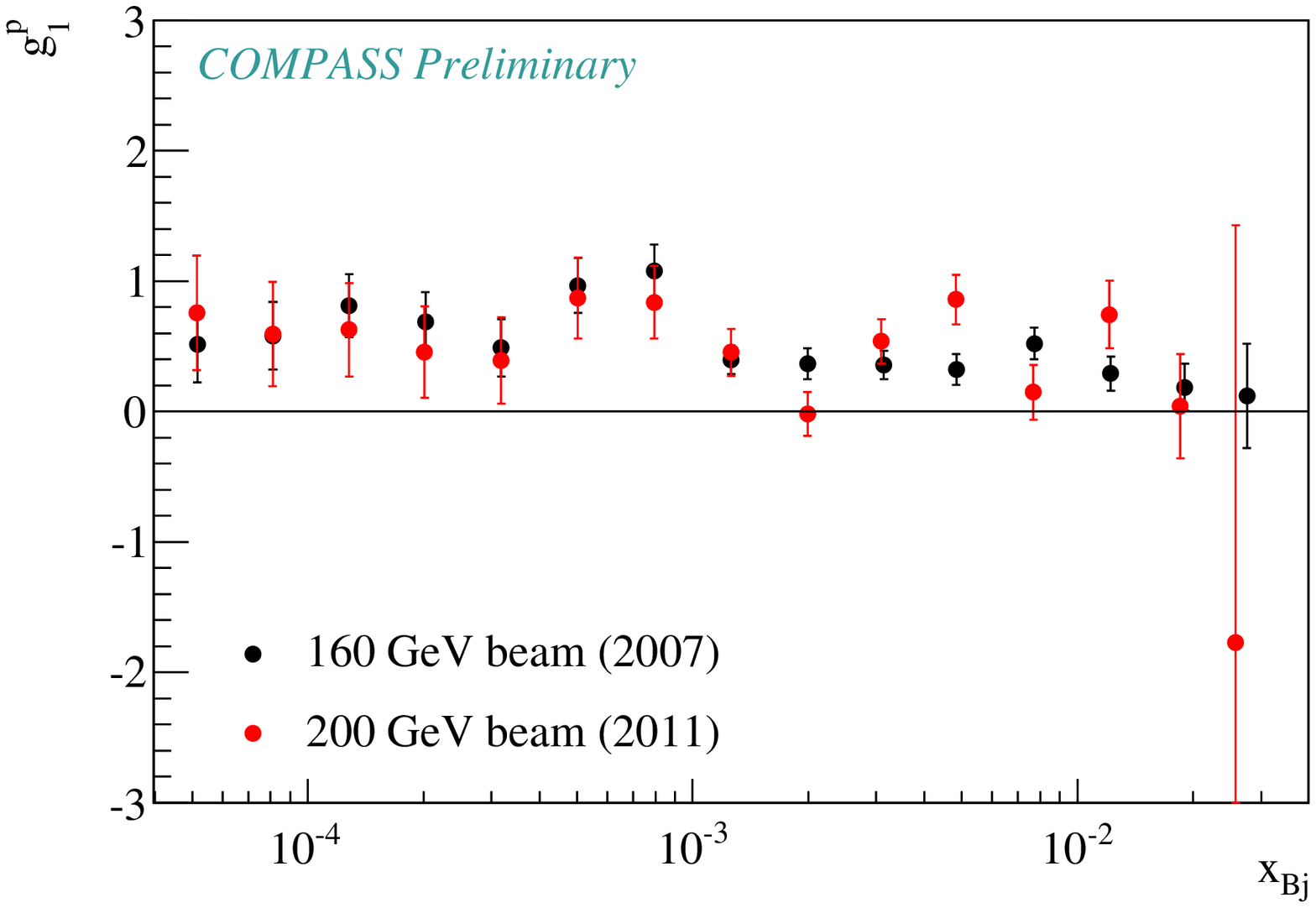}&
\includegraphics[width=0.5\textwidth]{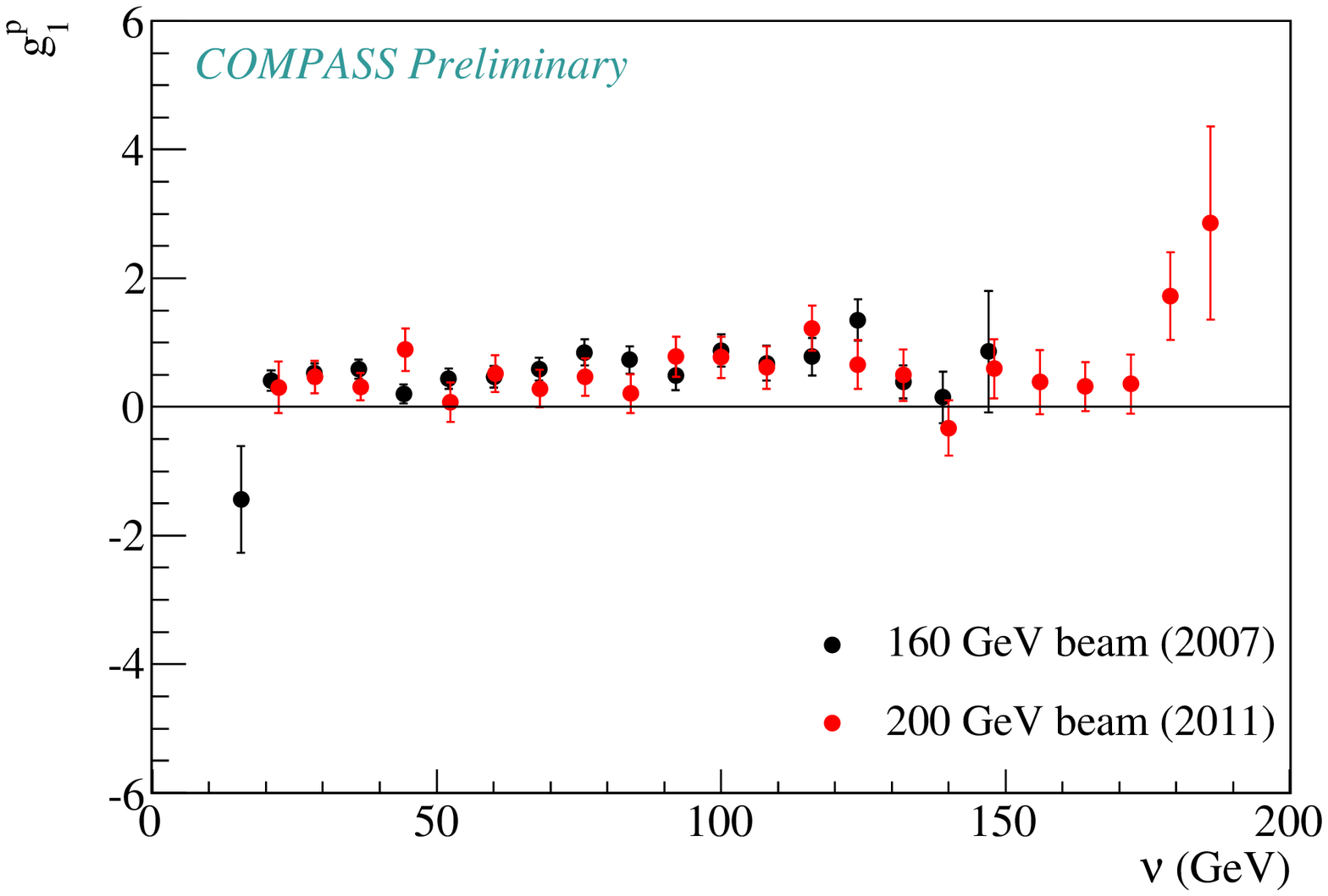}\\
    \textbf{(a)} & \textbf{(b)} \\
    \end{tabular}
      \caption{\footnotesize
    \textbf{(a)} Spin dependent structure function $g_1^p$ as a function of $x$, obtained from 2007 and 2011 data. The results for the two beam energies are compatible within errors. The systematic errors are expected to be smaller than the statistical errors. The asymmetries are mostly incompatible with zero and positive.
    \textbf{(b)} The same, but for $g_1^p$ as a function of $\nu$. 
  }
  \label{fig:g1}
\end{figure}

The spin dependent structure function of the proton, $g_1^p$, was also obtained independently in bins of $x$ and in bins of $\nu$, the virtual photon energy, according to
$$g_1^p=\frac{F_2^p(\langle x\rangle , \langle Q^2\rangle)}{2x[1+R(\langle x\rangle , \langle Q^2\rangle) ]}A_1^p,$$
where $F_2$ was obtained from a parameterisation from the SMC within its validity range~\cite{SMCgeneric}, or from a model otherwise (low $x$ and $Q^2$)~\cite{BKf2}; and $R$ was taken from data or, in the case of low $x$, from a parameterisation, as described in~\cite{COMPASSdeuteronQ2lt1}.
The preliminary results obtained for $g_1^p$ are presented in Fig.~\ref{fig:g1}. 

In both cases of $A_1^p$ and $g_1^p$ the preliminary results obtained with different beam energies of 160 GeV and 200 GeV are compatible within errors. No special dependence with $\nu$ is observed. Furthermore, the results are incompatible with zero and positive in the measured ranges. This is the first time that spin effects are observed at such low values of $x$. 
It will now be possible to extract a more precise non-singlet structure function $g_1^\text{NS}$ from the COMPASS results of $g_1^p$ and $g_1^d$ for $Q^2<1$ (GeV/$c$)$^2$, in order to compare it with theoretical predictions.


\end{document}